\newcommand{\dphi}{\Delta \phi}
\newcommand{\pion}{$\pi^{0}$}

\newcommand{\beq}{\begin{equation}}
\newcommand{\eeq}{\end{equation}}

\documentclass[final,1p,times]{elsarticle} 
\usepackage{graphicx} 
\usepackage{amssymb} 
\usepackage{amsthm} 
\usepackage{lineno} 
\usepackage{subfigure}
\usepackage{wrapfig}

\journal{Nuclear Physics A} 
\begin{document} 

\begin{frontmatter} 


\title{Probing High Parton Densities at Low-$x$ in d+Au Collisions at PHENIX Using the New Forward and Backward Muon Piston Calorimeters}

\author{Beau Meredith$^{a}$ for the PHENIX collaboration}

\address[a]{University of Illinois at Urbana Champaign, 
1110 W Green St, Urbana, IL, 61801, USA}

\begin{abstract} 
The new forward Muon Piston Calorimeters allow PHENIX to explore
low-$x$ parton distributions in d+Au collisions with hopes  of
observing gluon saturation.  We present a two-particle azimuthal
$\Delta \phi$ correlation measurement made between a mid-rapidity
particle ($|\eta_1| < 0.35$) and a forward $\pi^0$ ($3.1 < \eta_2 <
3.9$) wherein we compare correlation widths in d+Au to p+p and compute
$I_{dA}$.
\end{abstract} 

\end{frontmatter} 



\section{Introduction}
Deuteron-gold collisions at RHIC provide a system wherein one can
explore nuclear effects on initial-state parton densities in the
absence of final-state medium effects in heavy ion collisions.  RHIC
experiments have shown a suppression in nuclear modification factors
($R_{dA}$, $R_{cp}$) for $\sqrt{s_{NN}} = 200$ $GeV$ d+Au collisions
in the forward (deuteron) direction and an enhancement in the backward
(gold) direction [1-3].  Multiple theories exist that can explain
the observed suppression and enhancement, but a conclusive measurement
discriminating between the different mechanisms has yet to be carried
out.  Two new forward electromagnetic calorimeters (Muon Piston
Calorimeters or MPCs, $-3.7 < \eta < -3.1$, $3.1<\eta<3.9$) were
recently installed in the PHENIX experiment allowing study of parton
densities at low $x$.  The MPCs make it possible to measure nuclear
modification factors in the forward and backward directions as well as
azimuthal correlations of di-hadron pairs at different
pseudorapidities.  In these proceedings, we present the results of the
first correlation measurements made with the new MPCs.  The analysis
presented is based on the $\approx 80$ $nb^{-1}$ integrated luminosity
data sample of d+Au collisions at $\sqrt{s_{NN}} = 200$ $GeV$ taken at
RHIC in 2008.  The correlation measurements are especially interesting
because it is expected that they provide a strong test of gluon
saturation at low $x$ in the Au nucleus \cite{monojets}, \cite{cgc}.

Two-particle correlations have previously been measured at PHENIX in
$\sqrt{s} = 200$ $GeV$ d+Au and p+p collisions for two charged hadrons
($h^{\pm}$) detected by the central arm spectrometers
($|\eta_1,\eta_2| < 0.35$) \cite{ida_central_phenix} and for rapidity
separated particles where one particle is a $h^{\pm}$ in the central
spectrometer ($|\eta_1| < 0.35$) and the other particle is a
punch-through hadron \cite{rcp_phenix} detected in the forward
or backward muon spectrometer ($1.4 < |\eta_2| < 2.0$)
\cite{ida_phenix}.  Throughout these proceedings ``forward'' (``backward'')
refers to the direction of the deuteron (gold) beam.


The correlation measurement is performed with a $h^{\pm}$ or a
$\pi^0$ in the mid-rapidity detectors ($|\eta_1| < 0.35$) and
a $\pi^0$ detected in the forward MPC.  The forward $\pi^0$
serves to lower the $x$ at which we probe the gold nucleus and to
provide a larger rapidity gap ($\Delta \eta \approx 3.5$) than has been
observed previously ($|\Delta \eta| \approx 1.5$ for the mid-rapidity
$h^{\pm}$/punch-through hadron correlations).

\section{Muon Piston Calorimeters}
The MPCs are PbWO$_4$ electromagnetic calorimeters that add to
the PHENIX calorimeter acceptance in the interesting forward and
backward regions; one MPC is installed in the north muon piston hole
($3.1 <\eta < 3.9$) and the other in the south ($-3.7 < \eta < -3.1$),
and both cover an azimuthal angle of $2 \pi$.  The towers have lateral
dimensions of $2.2$ $cm$ $\times$ $2.2$ $cm$.  The MPCs are installed $220$ cm from
the nominal interaction point \cite{chiu}.

In any electromagnetic calorimeter, if the momentum of a $\pi^0$ is
sufficiently high, the energy from both decay photons will be
reconstructed as a single cluster.  For the MPC, this merging effect
dominates at and above $p_{tot} = 20$ $GeV/c$.  Hence, below $20$ $GeV/c$,
$\pi^0$s are identified by the invariant mass spectrum of all photon
pairs; while above $20$ $GeV/c$, $\pi^0$s are identified using single
clusters.

\begin{figure}[htbp]
\begin{center}
   \subfigure[]{}\includegraphics[scale=0.12]{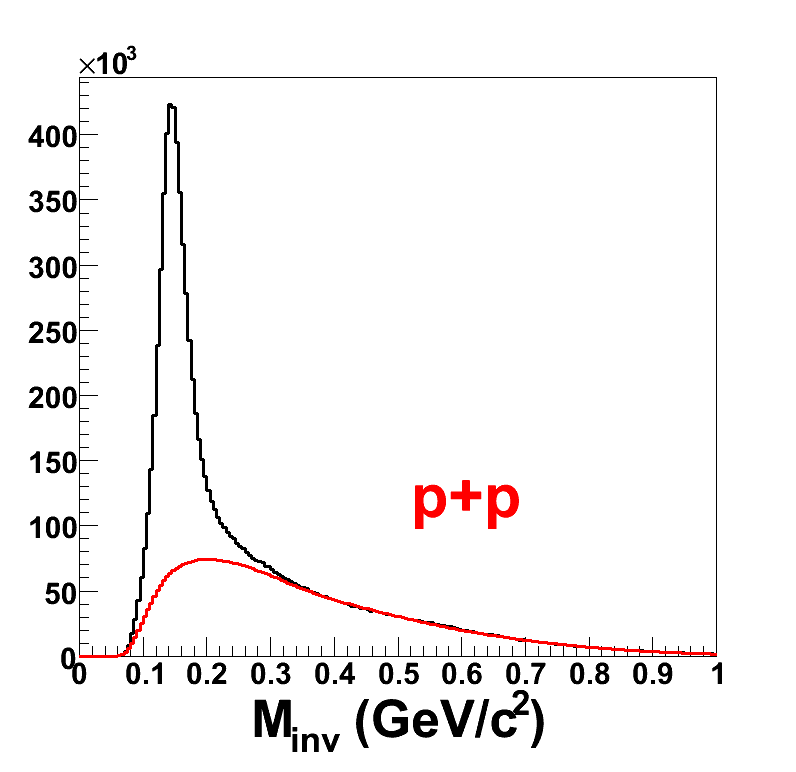}
   \subfigure[]{}\includegraphics[scale=0.12]{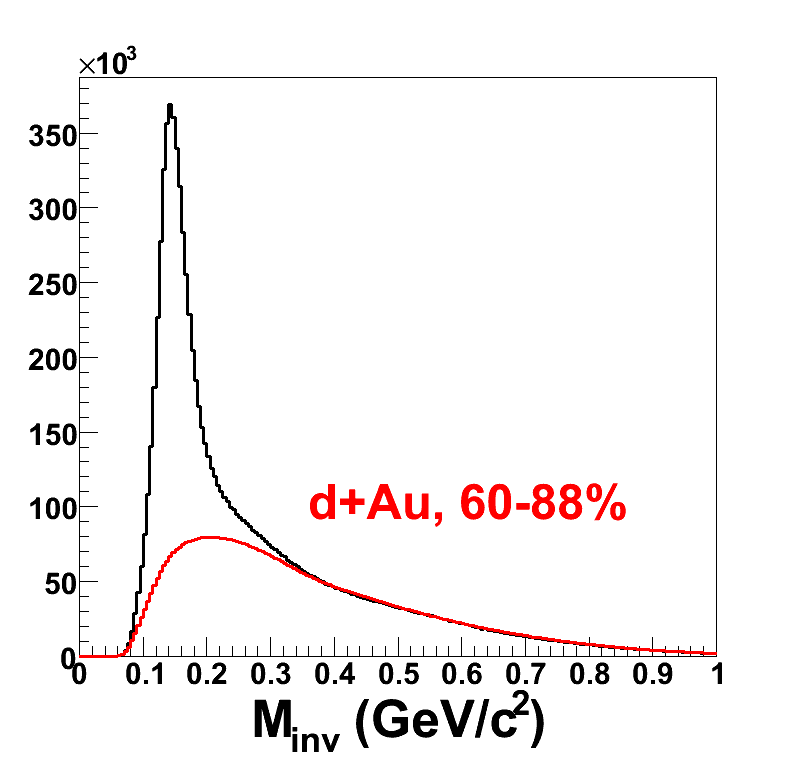}
   \subfigure[]{}\includegraphics[scale=0.12]{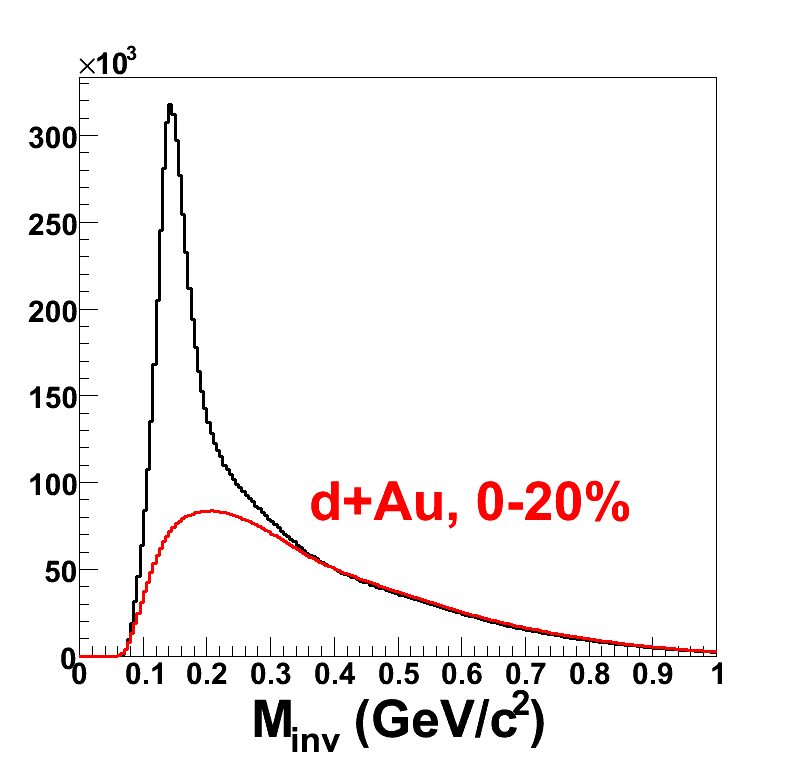}
\caption{North MPC invariant mass distributions of photon pairs (black) and the background distributions obtained using an event mixing method (red) for $0.68$ $GeV/c$ $< p_T < 0.91$ $GeV/c$ for  \textbf{(a)} p+p, \textbf{(b)} d+Au 60-88\% centrality bin, \textbf{(c)} d+Au 0-20\% centrality bin.}
\label{mpcmass}
\end{center}
\end{figure}

%
To identify $\pi^0$s from two photons, we use the following set of cuts:
$7$ $GeV$ $< E_{\gamma\gamma} <17$ $GeV$, a cluster separation cut of
$|\Delta r| > 3.5$ $cm$, an energy asymmetry cut on the two clusters
of $\alpha =\frac{E_2 - E_1}{E_2+E_1} < 0.6$.  To
identify $\pi^0$s using single clusters, we specify that $20$ $GeV$ $<
E_{\gamma\gamma} < 50$ $GeV$. Sample invariant mass plots are shown in Fig. \ref{mpcmass}.

%

%
%
%
%
%

\section{Data Analysis}


In the correlation analysis, the mid-rapidity particle is the trigger
particle, and the forward particle is the associate particle \cite{correlations}.  A peak at $\Delta\phi = 0$ is
not present in the $\Delta \phi$ distributions because the particles
are separated in rapidity by approximately 3.5 units, which is wider
than the width of the near side jet structure.  Hence only an away
side peak is expected at $\Delta\phi =
\pi$.

\begin{figure}[htbp]
\begin{center}
   \subfigure[]{}\includegraphics[scale=0.12]{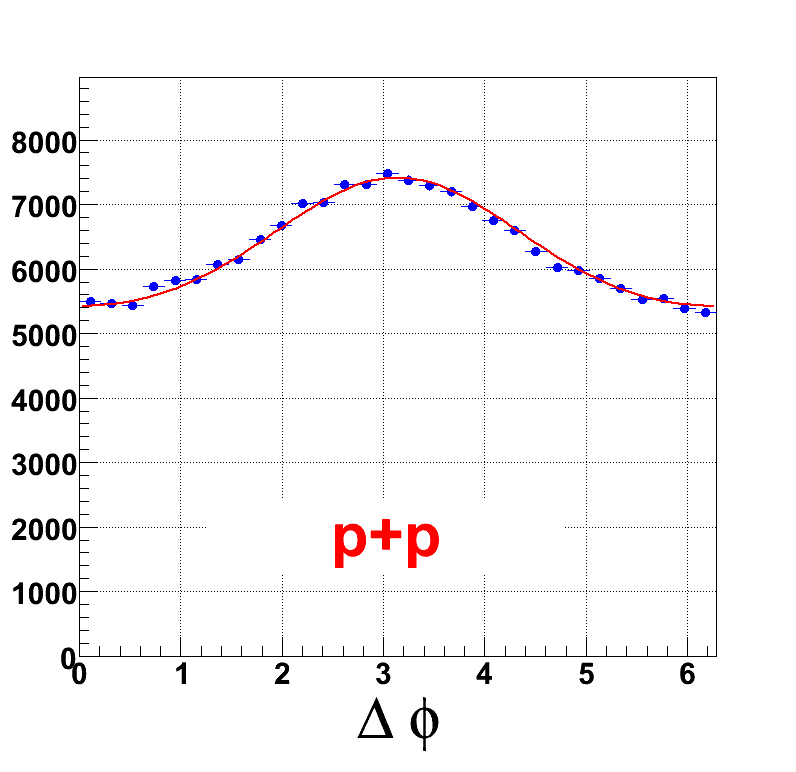}
   \subfigure[]{}\includegraphics[scale=0.12]{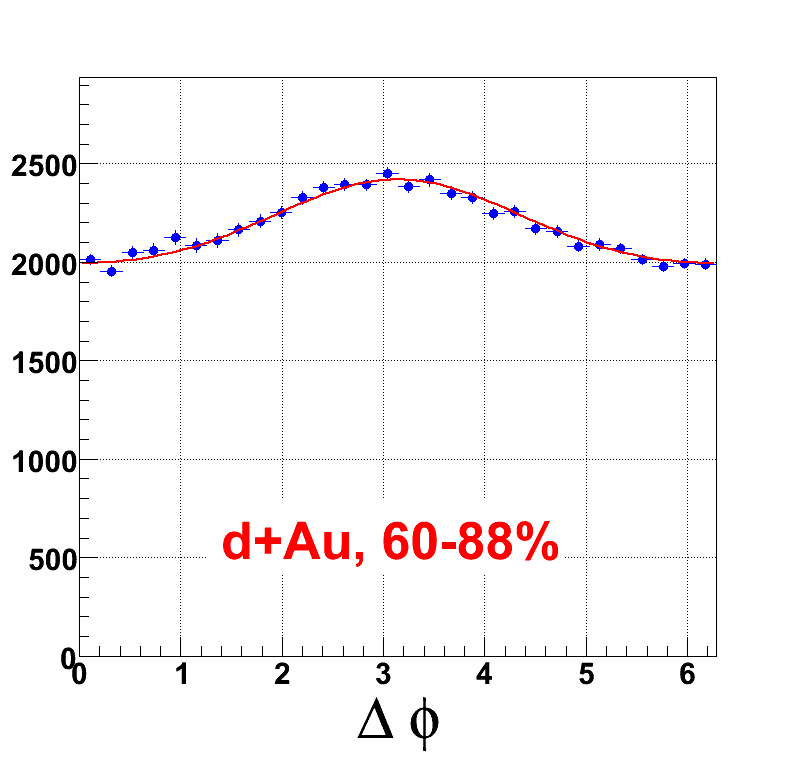}
   \subfigure[]{}\includegraphics[scale=0.12]{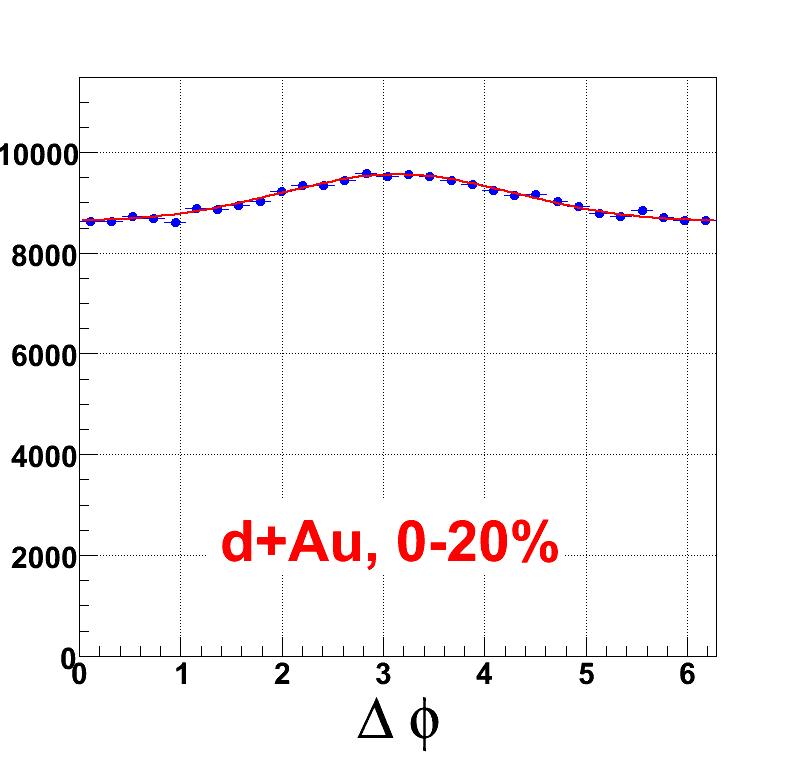}
\caption{Example $\Delta \phi$ correlation functions for trigger $\pi^0$ ($|\eta_1| < 0.35$) $2$ $GeV/c$ $< p_T < 3$ $GeV/c$, associate $\pi^0$ ($3.1 < \eta_2 < 3.9$) $0.45$ $GeV/c$ $< p_T < 0.68$ $GeV/c$ for  \textbf{(a)} p+p, \textbf{(b)} d+Au 60-88\% centrality bin, \textbf{(c)} d+Au 0-20\% centrality bin.}
\label{pi0corr}
\end{center}
\end{figure}

The correlation function, $CF(\Delta \phi)$, is the $\Delta \phi$
distribution of the two particles corrected for the nonuniform
detector acceptance ($acc(\Delta \phi)$ is the two-particle $\Delta \phi$
distribution where the particles are from different events), or
$CF(\Delta \phi) = \frac{1}{acc(\dphi)} \times
\frac{dN^{measured}(\dphi)}{d(\dphi)}$ \cite{correlations}.  Example
acceptance-corrected $\Delta \phi$ correlation functions from this
analysis are shown in Fig. \ref{pi0corr}.  The correlation functions
are fit as a Gaussian di-jet signal on top of a constant
background.  Two interesting quantities to compare d+Au with p+p are
the width of the Gaussian peak and the conditional (or per-trigger)
yield, $CY$.  The conditional yield is the efficiency-corrected
pair-yield of particles produced per trigger particle detected, or

\beq
CY = \frac{\int_{0}^{2 \pi} d(\Delta \phi) (CF(\Delta \phi) -
bg(\Delta \phi))}{N_{trig}\times\epsilon}
\eeq
where $N_{trig}$ is the number of trigger particles, $\epsilon$ is the
detection efficiency of the associate particle, and $bg(\dphi)$ is the
constant combinatorial background determined by fitting the
correlation function.  We then form a ratio of the $CY$s for d+Au and
p+p which is the nuclear modification factor $I_{dA}$:

\beq
I_{dA} = \frac{CY_{dA}}{CY_{pp}}
\eeq

 

\begin{figure}[htbp]
\begin{center}
   \subfigure[]{}\includegraphics[scale=0.2]{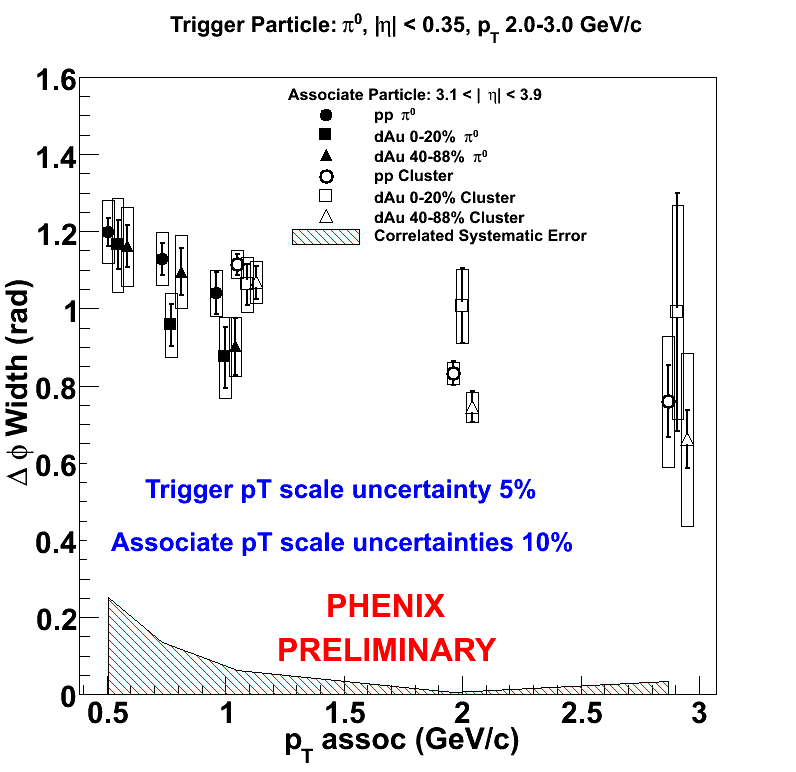}
   \subfigure[]{}\includegraphics[scale=0.2]{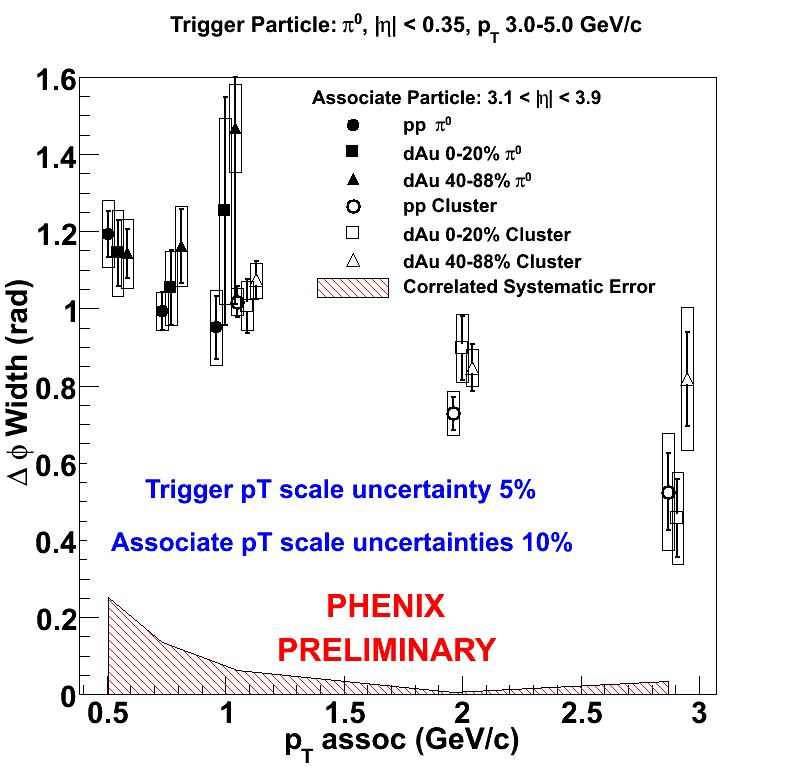}
\caption{$\Delta \phi$ width vs. $p_T$ for mid-rapidity \pion/forward MPC \pion correlations.  \textbf{(a)} mid-rapidity \pion $p_T = 2-3$ $GeV/c$. \textbf{(b)} mid-rapidity \pion $p_T = 3-5$ $GeV/c$.}
\label{w1}
\end{center}
\end{figure}

\clearpage

\section{Discussion}

The correlation widths are shown in Fig. \ref{w1}.  To extend the
$p_T$ range of the associate particle both the MPC
$\pi^0$ (two-photon identification) and MPC cluster widths are shown.
The widths decrease with increasing $p_T$ as expected from jet
fragmentation.  Within the precision of statistical and systematic
errors, little variation can be seen when comparing the correlation
widths for d+Au (central or peripheral) and p+p.  

\begin{wrapfigure}{r}{0.5\textwidth}
  \begin{center}
    \includegraphics[width=0.48\textwidth]{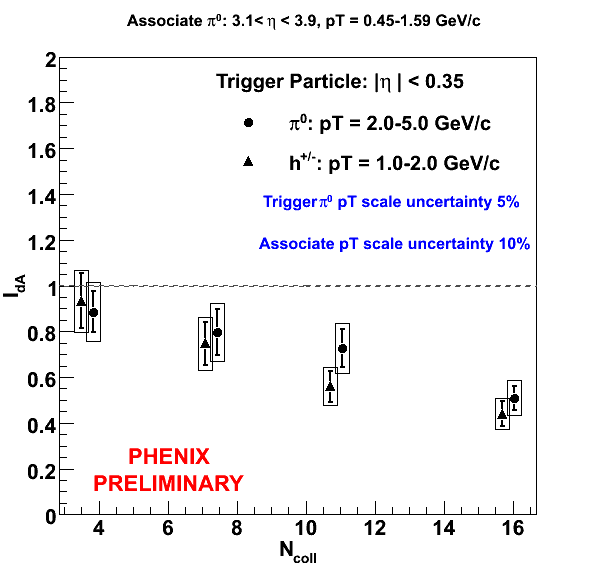}
  \end{center}
  \caption{\textbf{(a)} $I_{dA}$ vs. $N_{coll}$ for mid-rapidity hadron/forward MPC \pion correlations.}
\label{ida1}
\end{wrapfigure}

On the other hand, $I_{dAu}$ (see Fig. \ref{ida1}, formed only for MPC
$\pi^0$s and not clusters) shows a suppression with increasing
collision centrality for both species of trigger particles ($\pi^0$s,
$h^{\pm}$, $|\eta_1| < 0.35$) that is significant.

Multiple theories are being used to explain these results (especially
$I_{dA}$); the most notable are the Color Glass Condensate (CGC) model
for gluon saturation \cite{monojets, cgc} and pQCD based pictures such as non-leading twist shadowing used by Vitev
\cite{vitev}.  It will be interesting to see if the d+Au forward RHIC correlation results will be able to distinguish between the above models.  

This interesting result is an exciting start to a set of PHENIX
di-hadron correlations that will span different total rapidities (utilizing
all PHENIX detectors) and rapidity gaps, allowing studies of the effects over varying ranges of $x$.

%

%

\section*{Acknowledgments} 
This work is supported by NSF PHY 0601067 and by the Department of
Energy which operates RHIC and PHENIX.

\end{document}